\documentstyle[aps,prl,twocolumn,epsfig,floats]{revtex}

\tighten

\newcommand{\ii}{{\rm i}}
\newcommand{\de}{{\rm\,d}}
\newcommand{\e}{{\rm e}}

\newcommand{\im}{\,{\rm Im}\,}
\newcommand{\tr}{\mbox{Tr}\,}
                 
\newcommand{\st}{{\scriptscriptstyle T}}

\newcommand{\g}{\gamma}
\newcommand{\sig}{\sigma}
\newcommand{\eps}{\epsilon}
\newcommand{\nn}{\nonumber}
\newcommand{\ovl}{\overline}

\newcommand{\pslash}{\rlap{/} p}
\newcommand{\kslash}{\rlap{/} k}
\newcommand{\lslash}{\rlap{/} l}

\begin{document}
 
\twocolumn[\hsize\textwidth\columnwidth\hsize\csname
@twocolumnfalse\endcsname

\title{
\begin{flushright}
\begin{minipage}{4 cm}
\small
VUTH 01-03\\
\end{minipage}
\end{flushright}
The Collins fragmentation function: a simple model calculation}

\author{A.~Bacchetta, 
R.~Kundu, A.~Metz, P.~J.~Mulders}

\address{
Division of Physics and Astronomy, Faculty of Science, Free University \\
De Boelelaan 1081, NL-1081 HV Amsterdam, the Netherlands\\[2mm]}

\date{February 22, 2001}

\maketitle

\begin{abstract}
The Collins function belongs to the class of the so-called time-reversal odd 
fragmentation functions. Being chiral-odd as well, it can serve as an
important tool to observe the nucleon's transversity distribution 
in semi-inclusive DIS. 
Due to the possible presence of final state interactions, this function can be
non-zero, though this has never been demonstrated in an explicit model 
calculation.
We use a simple pseudoscalar coupling between pions and quarks to model the
fragmentation process and we show that the inclusion of one-loop corrections
generates a non-vanishing Collins function, therefore
giving support to its existence from the theoretical point of view. 
\end{abstract}


]


Our understanding of hadronic physics depends strongly on what
we know about the parton distribution and fragmentation
functions, which are universal, process-independent objects. 
While in the past several experiments provided us with considerable
information on the parton distributions, our knowledge of the fragmentation
functions is still rather limited. 

A lot of attention has been devoted to the class of the so-called 
{\it time-reversal odd} (T-odd) fragmentation functions, among which the 
Collins 
function $H_1^{\perp}$ \cite{Collins:1993kk,Levelt:1994np} 
is the most prominent example.
In contrast to a naive expectation, these functions can be non-vanishing 
because time-reversal invariance does not impose any constraint 
\cite{Collins:1993kk,Jaffe:1993xb} (in the literature they are sometimes 
referred to as {\it naive T-odd} functions
 \cite{Bianconi:2000cd}).
This is a consequence of possible 
{\it final state interactions} in the
fragmentation process, giving rise to a non-trivial phase (imaginary part)
\cite{Collins:1993kk}. We remark that final state interactions should not 
be considered exclusively
as reinteractions of the outgoing hadron with the rest of the jet,
as we will discuss in more detail later.

The Collins function describes the fragmentation of a transversely polarized 
quark into an unpolarized hadron (e.g.\ a pion), i.e. the process 
$q^{\ast} \to h X$.
The introduction of this function requires taking into account the quark's 
transverse momentum.
Although the Collins function in itself deserves attention 
towards understanding 
the physics governing the fragmentation process, it is particularly relevant
because in semi-inclusive DIS it can serve as the 
chiral-odd partner needed 
to access the transversity distribution of the nucleon,
$h_1$, which is  
chiral-odd as well.
The transversity is a crucial property of the nucleon, carrying information on
 its  spin structure 
 complementary to what we can deduce from helicity distributions.
Because of its chiral-odd nature, $h_1$ is suppressed in {\it inclusive} DIS,
and therefore it has remained essentially unexplored on the experimental side.
In the case of {\it one-particle inclusive} DIS, where one 
detects only a single hadron in the final state, the Collins function
provides 
the only possibility to probe $h_1$ at the leading twist level. 

Even if the Collins function can not be discarded in general 
on the basis of 
time-reversal invariance, it has been suspected to vanish since 
the effect arising
from final state interactions may average out \cite{Jaffe:1998pv}. 
Moreover, $H_1^{\perp}$ has been proven elusive to previous modeling 
attempts \cite{Bianconi:2000cd}. This suggested to turn the attention to the 
experimentally more challenging detection of {\it two} hadrons 
in a semi-inclusive 
measurement \cite{Collins:1994kq,Jaffe:1998pv},
as an alternative method to measure
the transversity. 
Until now, no {\em ab initio} calculation has ever displayed the possibility of
generating a non-zero Collins function in the framework of a simple,
time-reversal invariant model.

On the experimental side, the HERMES collaboration reported the first
observation of a single-spin asymmetry in semi-inclusive pion production 
\cite{Airapetian:2000tv}, which could be interpreted as arising from the
contribution of the Collins function \cite{Oganessyan:1998ma}. Analogously,
large single-spin asymmetries in the process $p^{\uparrow}p\rightarrow \pi X$
\cite{Bravar:1996ki} could also be explained by means of the Collins 
function \cite{Anselmino:1999pw}.
The situation is far from clear, but further investigations are
among the priorities of the HERMES~\cite{hermes}, COMPASS~\cite{compass}
and possibly {eRHIC} programs.

In these circumstances, producing a non-zero Collins function in a simple yet 
consistent model is an interesting and relevant result, which we are 
going to present in this letter. 
To this end we describe the fragmentation process by a pseudoscalar coupling
between quarks and pions in the one-loop approximation.
A similar calculation has been suggested earlier \cite{Suzuki:2000pe}, 
but never 
carried out explicitly.
For what concerns possible phenomenological applications, we do not pretend 
our calculation 
to be realistic, but we think that it can shed light on the identity of
T-odd fragmentation functions and can conclusively affirm that there are 
no theoretical reasons to believe the Collins function to vanish.


Considering the fragmentation process $q^{\ast}(k) \to \pi(p) X$,
 we define the Collins function, 
which depends on the longitudinal momentum fraction $z$ of the pion and
the transverse momentum $k_{\st}$ of the quark, as \cite{Mulders:1996dh} 
\footnote{Note that this definition of $H_1^{\perp}$ slightly differs from the
original one given by Collins \cite{Collins:1994kq}.}
\begin{equation} 
\frac{\eps_{\st}^{ij} k_{\st\,j}}{m_{\pi}}
				\, H_1^{\perp}(z,k^2_{\st}) =
	 \left. \frac{1}{4z} \int \de k^+ \;
              \tr[ \Delta (k,p) \ii \sig^{i-}\g_5]
	\right|_{k^-=\frac{p^-}{z}},  
\label{e:col1}
\end{equation}
with $m_{\pi}$ denoting the pion mass and $\eps_{\st}^{ij} \equiv \eps^{ij-+}$
(we specify the plus and minus lightcone components of a generic 4-vector 
$a^{\mu}$ 
according to $a^{\pm} \equiv (a^0 \pm a^3)/\sqrt{2}$).
The correlation function $\Delta(k,p)$ in Eq.~(\ref{e:col1}), omitting gauge
links, takes the form
\begin{equation} 
\Delta(k,p)=\sum_X\, \int
        \frac{\de^{4}\xi}{(2\pi)^{4}}\e^{+\ii k \cdot \xi}
       \langle 0|
\,\psi(\xi)|\pi, X\rangle \langle \pi, X|
             \ovl{\psi}(0)|0\rangle.    
\label{e:delta}
\end{equation} 


To describe the matrix elements in the correlation function, we use a 
pseudoscalar coupling between quarks and pions 
given by the interaction Lagrangian 
\begin{equation}
{\cal L}_I(x) =  \ii g \,\ovl{q} (x) \g_5 q(x) \pi(x) \,,
\label{e:lagrangian}
\end{equation}
which is in the spirit of the Manohar--Georgi model \cite{Manohar:1984md}.
This is clearly an oversimplified approach to the fragmentation 
$q^{\ast} \to \pi X$
but the model contains the essential elements required for our discussion. 
In particular, it is time-reversal invariant.
One could also perform the calculation in a chirally invariant model by
including a scalar $\sigma$ field as well as taking quark flavors properly 
into account.
In our view, this is certainly necessary to improve the result obtained here
for a better description of the phenomenology,
but the essential result of a non-zero $H_1^{\perp}$ is already evident
without going to the complications arising from the inclusion of the $\sigma$
particle and the discussion of various flavors.

\begin{figure}
        \centering
        \epsfig{figure=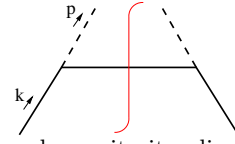,width=3cm}
        \caption{Lowest-order unitarity diagram describing the fragmentation 
		of a quark into a pion.} 
	\label{f:born}
        \end{figure}

Using the Lagrangian in Eq.~(\ref{e:lagrangian}), 
at tree level the fragmentation of a quark
is modeled through the process $q^{\ast} \to \pi q$.
The corresponding correlation function can be represented by the unitarity
diagram in Fig.~\ref{f:born} and reads explicitly
\begin{eqnarray} 
\Delta_{(0)}(k,p)&=&-\frac{g^2}{(2 \pi)^4}\,\frac{(\kslash + m)}{k^2 -m^2}\,
 \g_5\, (\kslash - \pslash +m)\, \g_5\,
\frac{(\kslash + m)}{k^2 -m^2} \nn \\
&&\mbox{}\times 2 \pi\,\delta((k-p)^2 -m^2) \,,
\end{eqnarray} 
where $m$ represents the mass of a constituent quark.
Using this correlation function, the unpolarized fragmentation function
\begin{equation} 
D_1(z,k^2_{\st}) =
\left. \frac{1}{4z} \int \de k^+ \;
              \tr[ \Delta (k,p) \g^-] \right|_{k^-=\frac{p^-}{z}} 
\end{equation} 
in our model at tree level becomes
\begin{equation} 
D_1(z,k^2_{\st}) = \frac{1}{z}\frac{g^2}{16 \pi^3}\,
	\frac{k_{\st}^2 +m^2}{(k_{\st}^2 +m^2+\frac{1-z}{z^2}m_{\pi}^2)^2} \,.
\label{e:unpolfrag}
\end{equation}
In the case of $m_{\pi} = 0$ we recover the result already obtained by
Collins~\cite{Collins:1993kk}. 
Contrary to $D_1$, the Collins function at tree level is zero. 
This is not surprising because at tree level no
{\it final state interaction} appears 
in the fragmentation process, which is supposed 
to be the origin of the T-odd fragmentation functions 
like $H_1^\perp$. 
The situation changes when we proceed to one-loop corrections, 
as we explicitly show in the following.

A consistent one-loop calculation 
of the fragmentation process requires the evaluation of
all the diagrams shown in Fig. 2. 
In the actual calculation only the asymmetric diagrams (a--d) contribute.
The contributions from the diagrams (e), (g) and (h) vanish individually
when projecting out the Collins function, while the contributions of
diagram (f) and
its hermitian conjugate cancel each other. 

It should be noted that not all the asymmetric (interference-type)
diagrams contribute. For example, the diagram (f) and its
hermitian conjugate represent the interference between two different tree
level amplitudes describing the process
$q^{\ast}\rightarrow\pi \pi q$. However,
their contributions to the Collins function cancel each other since
the involved interfering amplitudes are purely {\it real}. 
In calculating $H_1^{\perp}$, such a cancellation between contributions 
coming from a diagram
and its hermitian conjugate takes place whenever the involved amplitudes 
are purely real and is a model independent feature.

In our computation, the relevant components to be included are the
self-energy correction for
diagrams (a) and (b), and the vertex correction for
diagrams (c) and (d).
 These corrections are sketched in
Fig.~\ref{f:sigmagamma} and can be analytically expressed as  
\begin{eqnarray}
-\ii \Sigma (k) & = & - g^2 \int \frac{\de^4 l}{(2 \pi)^4}\,
      \frac{\lslash - \kslash + m}{[(k-l)^2 - m^2]\,[l^2 - m_{\pi}^2]} \,, \\
g \g_5 \Gamma(k,p) & = & - \ii g^3 \g_5 \int \frac{\de^4 l}{(2\pi)^4} \,
      \frac{(\kslash - \pslash - \lslash + m)}{[(k - p - l)^2 - m^2]} \nn \\
&& \mbox{}\times \frac{(\lslash - \kslash + m)}
      {[(l - k)^2 - m^2]
	 [l^2 - m_{\pi}^2]} \,.
\end{eqnarray} 
The functions $\Sigma (k)$ and $\Gamma (k,p)$ can be parametrized as
\begin{eqnarray} 
\Sigma (k) &=& A\,\kslash + B\, m \,, \label{e:sig}\\
\Gamma (k,p) &=& C + D\,\pslash + E\,\kslash + F\,\pslash \, \kslash .
\label{e:ver}
\end{eqnarray}  
The real parts of the functions $A$, $B$, $C$ etc. 
are UV-divergent and require in principle a proper renormalization.
Though our model is renormalizable, we do not have to deal with this question 
at all, since only the imaginary parts of the loop diagrams
will turn out to be important.

	\begin{figure}
	\centering
	\epsfig{figure=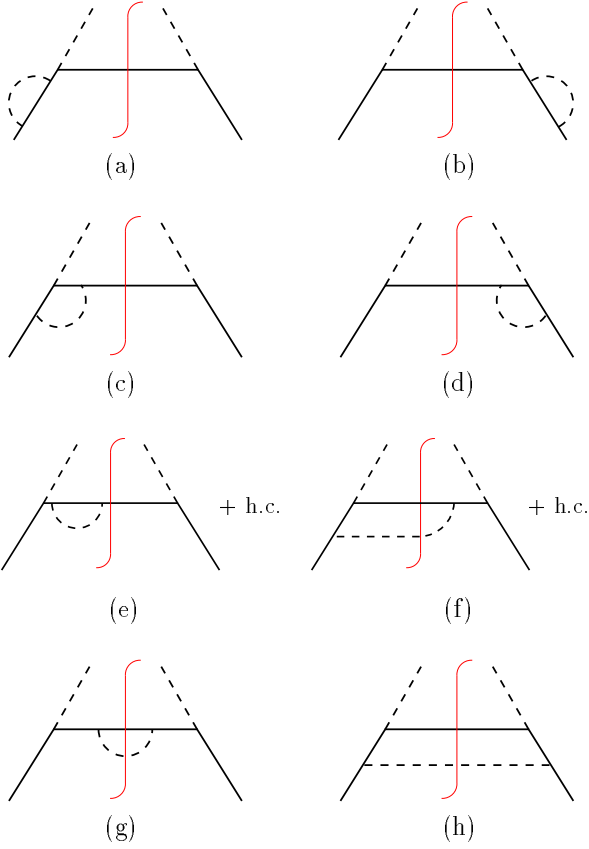,width=8cm}
        \caption{One-loop corrections to the fragmentation of a quark
		into a pion.}
        \label{f:1loop}
        \end{figure}

The contributions to the 
correlation function generated by the diagrams (a) and (c) are given by:
\begin{eqnarray} 
\lefteqn{\Delta_{(1)}^{(a)}(k,p) = -\frac{g^2}{(2 \pi)^4}\,\frac{(\kslash + m)}
  {k^2 -m^2} \, \g_5 \, (\kslash - \pslash +m)\,\g_5} \nn \\
&& \quad \mbox{}\times\frac{(\kslash + m)}{k^2 -m^2}\, \Sigma(k) \, 
        \frac{(\kslash + m)}{k^2 -m^2} \;2 \pi\;\delta((k-p)^2 -m^2) \,, \\
\lefteqn{\Delta_{(1)}^{(c)}(k,p) = -\frac{g^2}{(2 \pi)^4}\,\frac{(\kslash + m)}
  {k^2 -m^2} \, \g_5 \, (\kslash - \pslash +m)\, \g_5} \nn \\
&& \quad \mbox{}\times  \Gamma(k,p) \, \frac{(\kslash + m)}{k^2 -m^2} \;
        2 \pi \; \delta((k-p)^2 -m^2). 
\end{eqnarray} 
The contributions from diagrams (b) and (d) follow from the hermiticity
condition:
$\Delta_{(1)}^{(b)}(k,p)=\gamma^0\Delta_{(1)}^{(a) \dagger}(k,p)\gamma^0$,
$\Delta_{(1)}^{(d)}(k,p)=\gamma^0\Delta_{(1)}^{(c) \dagger}(k,p)\gamma^0$.

	\begin{figure}
	\centering
	\epsfig{figure=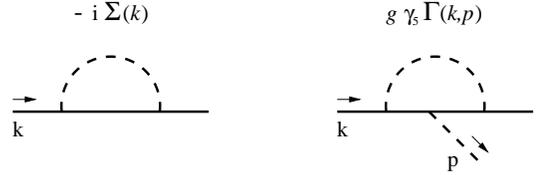,width=7cm}\\
        \caption{One-loop self-energy and vertex corrections.}
        \label{f:sigmagamma}
        \end{figure}

Summing the contributions of the four diagrams and inserting the resulting
correlation function in Eq.~(\ref{e:col1}), we obtain the result
\begin{eqnarray} 
\lefteqn{H_1^{\perp}(z,k^2_{\st})}\nn \\
& =& \frac{g^2\,m_{\pi}}{4 \pi^3} \,\frac{p^-}{z} \int \de k^+
	\delta((k-p)^2 - m^2) \nn \\
&&	\mbox{}\times
	\left( \frac{m\, \im(A+B)}{(k^2 -m^2)^2}
	+ \frac{\im (D+E+mF)}{(k^2 -m^2)}\right) 
	\biggr|_{k^-=\frac{p^-}{z}} \nn \\
&=&	\frac{g^2\,m_{\pi}}{8 \pi^3}\frac{1}{1-z}
	\left( \frac{m\, \im(A+B)}{(k^2 -m^2)^2}\right. \nn \\
&&	\left.\mbox{}+ \frac{\im (D+E+mF)}{(k^2 -m^2)}\right) 
	\biggr|_{k^2=\frac{z}{1-z}k_{\st}^2
	+\frac{m^2}{1-z} +\frac{m_{\pi}^2}{z}}\,.
\label{e:h1}
\end{eqnarray} 
Thus the actual value of the Collins function in this model depends only on the
imaginary parts of the coefficients defined in Eqs.~(\ref{e:sig}--\ref{e:ver}).
The lack of an imaginary component in these coefficients would inevitably
result in a vanishing Collins function.
We can compute the imaginary parts
by applying the Cutkosky rule to the self-energy
and vertex diagram of
Fig.~\ref{f:sigmagamma}. In this way, as mentioned before, 
we can avoid the issues related 
to renormalization, which affect only the real parts of the diagrams. 
Explicit calculation leads to
\begin{eqnarray}
\im(A+B) &=& \frac{g^2}{16 \pi^2}\left(1-\frac{m^2- m_{\pi}^2}{k^2}\right)I_1
			\,,
\label{e:first}\\
\im(D+E+m\,F) &=&-\frac{g^2}{8 \pi^2}\, m\, 
	\frac{k^2 -m^2 +m_{\pi}^2}{\lambda(k^2,m^2,m_{\pi}^2)} \nn \\
&&\mbox{}\times \left[I_1 + (k^2 - m^2 -2m_{\pi}^2) I_2\right],
\end{eqnarray} 
where we have introduced the so-called K\"allen function, 
$\lambda(k^2,m^2,m_{\pi}^2)=[k^2 -(m+m_{\pi})^2][k^2 -(m-m_{\pi})^2]$,
and the factors
\begin{eqnarray}
I_1 & = & \int \de^4 l \; \delta (l^2 - m_{\pi}^2)\, \delta ((k-l)^2-m^2) \nn \\
 & = & \frac{\pi}{2 k^2} \sqrt{ \lambda(k^2,m^2,m_{\pi}^2)}
	\;\theta(k^2 -(m+m_{\pi})^2) \,, \\
I_2 & = & \int \de^4 l \; \frac{\delta(l^2 - m_{\pi}^2)\,\delta((k -l)^2-m^2)}
       {(k - p - l)^2 - m^2} \nn \\
 & = & -\frac{\pi}{2 \sqrt{\lambda(k^2,m^2,m_{\pi}^2)}}
	\ln{\left(1+\frac{\lambda(k^2,m^2,m_{\pi}^2)}{k^2m^2 -(m^2-m_{\pi}^2)^2 }
	\right)} \nn \\ 
&&	\mbox{}\times 
	\theta(k^2 -(m+m_{\pi})^2).
\label{e:last}
\end{eqnarray} 
These integrals are finite and 
vanish below the threshold of quark-pion production, where the self-energy and
vertex diagrams do not possess any imaginary part.

Thus Eq.~(\ref{e:h1}) in combination with Eqs.~(\ref{e:first}--\ref{e:last})
gives an explicit non-vanishing result for the Collins function.

We point out that Collins suggested the idea of dressing the quark propagator
as a possible mechanism to produce a non-zero
$H_1^{\perp}$~\cite{Collins:1993kk}.  Here, we have not
only supported this conjecture by means of an explicit one-loop 
calculation, but
we have also shown that the contributions due to the self-energy and the
vertex corrections do not cancel each other.

\begin{figure}
	\centering
	\epsfig{figure=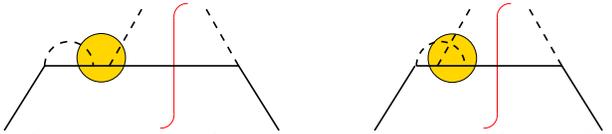,width=8cm}\\
        \caption{Interpretation of the one-loop corrections as final state
	interaction.} 
        \label{f:FSI}
        \end{figure}

To avoid possible confusions, we would like to make a few remarks on the issue
of the final state interaction. 
One-loop corrections in our model can be viewed as containing a
{\it specific example} of final state interaction, 
where the pion, after being emitted and before being detected, rescatters off
the quark through the direct and crossed channels
(see Fig.~\ref{f:FSI}).  
This interpretation is possible because the outgoing hadron and the hadron
appearing in the loop are the same. In general, however, 
it is too restrictive to treat
final  state interactions exclusively as a reinteraction of the outgoing
hadrons. In fact, in an interference-type cut-diagram, {\it any} kind of final
state interaction leading to an imaginary part in the amplitude of the process
$q^{\ast} \to h X$ generates a contribution to the Collins function. In our
case, for instance, it is sufficient to employ a different particle in
the loop to generate a final state interaction which is not a rescattering of
the outgoing hadron.

As a final remark, we would like to mention that the model we discussed
could be applied to describe the quark
distribution inside a pion in an attempt 
to generate T-odd distribution
functions~\cite{Sivers:1990cc,Boer:1998nt}. However, one can readily
see that, 
because of 
the different kinematical 
conditions, self-energy and vertex diagrams 
do not acquire an imaginary part, 
which is essential for producing a non-zero T-odd function.


In conclusion, we have shown that the Collins function can be generated using
a simple pseudoscalar coupling between quarks and pions to model the 
fragmentation process.
The Collins function turns out to be zero at the tree level due to the absence 
of any final state interaction which is at the origin of its existence, 
but renders itself at the one-loop level.
The calculation performed here suggests that the Collins function,
being non-zero already in a simple model, is very unlikely
to vanish in reality.
It is
therefore a worthwhile task to pursue the measurement of $H_1^{\perp}$ in DIS
and $e^{+}e^{-}$ annihilation. Moreover, our result supports the idea of
using {\it one-particle inclusive} DIS as a promising process to investigate
the transversity distribution of the nucleon.


We would like to thank E.~Leader and D.~Boer for useful discussions.
This work is part of the research program of the Dutch
Foundation for Fundamental Research on Matter (FOM) and it is partially 
funded by the European Commission IHP program under
contract HPRN-CT-2000-00130.


\end{document}